# Pressure Induced Metallization of $BaMn_2As_2$

A. T. Satya, Awadhesh Mani, A. Arulraj, N. V. Chandra Shekar, K.Vinod, C. S. Sundar and A. Bharathi*

Condensed Matter Physics Division, Materials Science Group,
Indira Gandhi Centre for Atomic Research,
Kalpakkam - 603 102 India

## Abstract

Temperature and pressure dependent electrical resistivity ($\rho$(T,P)) studies have been performed on $BaMn_2As_2$ single crystal in the 4.2 to 300 K range upto of 8.2 GPa, in order to investigate the evolution of its ground state properties. $\rho$(T) data show negative co-efficients of resistivity upto a pressure 3.2 GPa. At a pressure of ~4.5 GPa an insulator to metal transition is seen to occur at ~36 K, as indicated by a change in the temperature co-efficient of the $\rho$(T). However at a pressure of ~5.8 GPa the sample is metallic in the entire temperature range. XRD studies performed as a function of pressure, at room temperature, also shows an anomaly in the pressure versus volume curve around P ~ 5 GPa, without a change in crystal structure, indicative of an electronic transition, in support of the resistivity results. In addition to metallisation, a clear precipitous drop in $\rho$(T ) is seen at ~17 K for P $\geq$5.8 GPa.

*bharathi@igcar.gov.in

The hectic pace of research fuelled by the discovery of superconductivity in the FeAs compounds both by electron and hole doping [1], by the application of pressure [2,3,4] and iso-electronic substitutions [2,5], has provided impetus to search for superconductivity in compounds with similar structures[1]. Of these classes of compounds $BaMn_2As_2$ is interesting as it occurs in an anti-ferromagnetic insulating ground state [6,7,8]. The Mn atoms in this structure order antiferro-magnetically in a checker board fashion termed as G-type anti-ferromagnet, with the spins aligned collinear to the c-axis[8]. The Neel temperature is 625 K [8]. The local moment at the Mn site is large of ~3.88$\mu_B$[8] contrary to the small moment of 0.9 $\mu_B$ per Fe atom [1] seen in the spin density wave ground state of $BaFe_2As_2$. Resistivity and specific heat measurements [9] on flux grown single crystals indicate that the compound is a small band gap semiconductor. Band structure calculations also indicate that $BaMn_2As_2$ is anti-ferromagnetic and has a semi conducting band gap of 0.2eV [9]. It has been suggested that doping $BaMn_2As_2$ with carriers can induce large anti-ferromagnetic fluctuations in the metallic compound and can lead to exotic superconducting ground state, with a larger $T_C$, as compared to that seen in $BaFe_2As_2$ [1]. With this in view, several transition metals were doped at the Mn site, but the solubility turned out to be negligible and no metallicity could be obtained for substitutions with Cr, Fe, Co, Ni, Cu, Ru, Rh, Pd, Re, and Pt at the Mn site nor Sb at As site [10]. Several investigations [10,11] of $Ba(Fe_{1-x}Mn_x)_2As_2$ have been carried out, which indicate the occurrence of a miscibility gap [10] in the x=0.12 to x=1 composition range, the destruction of stripe like magnetic order[11] at x=0.102 etc., but the resistivity versus temperature in these Fe rich samples for (x~0.1) are seen to have an insulating behaviour [10,11]. Here, we provide a definitive evidence for metallization of single crystals of $BaMn_2As_2$ under the application of an external pressure of ~5 GPa using low temperature resistivity measurements in a high pressure cell. Also associated with metallization at 5.8GPa is a sharp fall in resistivity, seen at ~17 K indicating a possibility of superconductivity in this compound. Further, from the measurements of room temperature lattice parameter variations as a function of pressure in the $BaMn_2As_2$ crystals, we show an anomalous change in the pressure volume curve at the pressure at which metallization is seen in the resistivity data.

$BaMn_2As_2$ single crystals were synthesized using Ba chunks and pre-prepared MnAs powder precursors with out using any flux. MnAs was prepared by heating intimate mixtures of Mn powder and As powders in quartz crucibles in a stainless steel chamber that could be locked under a Ar pressure of 30 Bar[12], very similar to the procedure employed for the synthesis of FeAs powders. Stoichiometric amounts of Ba chunks and MnAs powder were taken in an alumina crucible and sealed in an evacuated quartz tube. The samples were then heated to $1200^0$C at a rate of $50^0$C per hour and held there for 24 hours, and then slow cooled to $800^0$C at a rate of $1.5^0$C per hour followed by a fast cooling rate of $50^0$ C up to room temperature. Shiny crystals with flat plate like morphology were formed with an average size of 1mm x 1mm x 0.4mm (see Fig.1a). The composition of the crystals, determined by Energy dispersive X-ray (EDX) attachment of an Scanning Electron Microscope, was consistent with $BaMn_2As_2$ stoichiometry, and also indicated the absence of any other elements (see Fig.1c), implying no contamination from the crucibles. The ambient XRD pattern of the powdered crystals were measured using Mo K$\alpha$, in the parallel beam transmission geometry and is shown in Fig. 2. It is evident from the figure that $BaMn_2As_2$, is phase pure and the small impurity lines could arise from inadvertent contamination from grinding. The XRD pattern could be indexed to tetragonal structure, space group I4/mmm, with lattice parameter values a = 4.1677(5)Å, c = 13.4686 (4) Å and V = 233.94 (7) Å$^3$, in good agreement with the previous results [7,8,9].

High pressure X-ray diffraction (HPXRD) experiments were carried out at room temperature using a Mao-Bell type diamond anvil cell (DAC) in an angle dispersive mode upto 10 GPa.

The sample in the powder form was loaded into a 200 μm diameter hole drilled in Stainless Steel gasket. A mixture of methanol, ethanol and water in the volume ratio 16:3:1 was used as pressure transmitting medium. The incident Mo X-ray beam obtained from a Rigaku ULTRAX (18kW) rotating anode X-ray generator was monochromatised with graphite monochromator. An image plate based mardtb345 diffractometer was used. The overall resolution of the diffractometer system is $\delta d/d \sim 0.001$. The Mao-Bell type DAC was fitted to the diffractometer and the sample to detector distance was calibrated using a standard specimen like $LaB_6$. The equation of state of silver was used as a parameter for pressure calibration. 2D image data from the Mar345 IP detector was converted to 1D Intensity versus 2 theta plots using the Fit2D program [13]. A sample XRD pattern carried out in the DAC at 0.34 GPa is diplayed in the inset of Fig.2 Each data set could be indexed with a tetragonal lattice with space group I4/mmm. Accurate lattice parameters were obtained by Rietveld refinements employing the GSAS program[14]. The detector to sample distances for ambient and high pressures were different leading to slight changes in resolution seen in the figure.

High pressure resistivity measurements as a function of temperature in the 4.2 K to 300 K temperature range, on a single crystal sample were carried out in a home built, opposed anvil pressure-locked cell. Steatite was used as the pressure transmitting medium and pyrophyllite washers were used as the gasket. The high pressure measurements on $BaMn_2As_2$ samples were carried out by maintaining similar experimental conditions to those used during in pressure calibration. The internal pressure of the cell was pre-calibrated by measuring the shift in the superconducting transition temperature of lead with respect to the applied load prior to mounting the samples. An error of ~0.2 GPa in the reported pressure can be expected. Further details about the sample assembly and measurements on different samples can be found in [4,15,16].

Figure 3 shows the temperature variation of resistance from 4.2 K to 300 K for representative pressures between 0 to 8.2 GPa. At ambient pressure (cf. Fig.3a) the resistance increases with decrease of temperature characteristic of a semiconductor. We could not measure the resistance below 30 K at ambient pressure as the value exceeded the measurement limit of the instrument. It must be emphasized that no metallic regime was observed in the R(T) data at ambient pressure, in contrast with the observation of metallic behaviour seen above 100 K for single crystals of $BaMn_2As_2$ synthesized by MnAs flux [7,9] and above 250 K for Sn flux grown $BaMn_2As_2$ [7]. These metallic behaviours were attributed to very small amount of inclusions of MnAs or Sn into the lattice [7,9]. At a pressure of 0.8GPa, the R(T) again showed insulating behaviour in the 70 K to 300 K temperature range, although a significant drop in resistance with respect to the ambient pressure data was seen at low temperature. With a further increase in applied pressures to 2.1GPa and 3.4 GPa (see Fig.3b and 3c), the resistance at low temperature decreases, with a discernable two step variation of R(T). With a further increase in pressure to 4.3 GPa, the variation R(T) shows a change in sign at 36 K, indicating a metallic behaviour at low temperature. Thus at this pressure the crystal shows an insulator to metal transition as a function of temperature. At 5.8 GPa, R(T) shows a metallic behaviour in the whole temperature range of 4.2 K to 300 K. With further increase of pressure from 5.8GPa to 8.2 GPa, the resistance further reduces at all temperatures, remaining metallic in the temperature range 4.2 K to 300 K. The variation in the resistance as function of pressure, are plotted for 30 K, 40 K and 290 K in Fig. 4a. The resistance at 30 K is seen to reduce by 7 orders of magnitude, at 40 K by 5 orders of magnitude and at 290 K by 3 orders of magnitude with increase in pressure to 8 GPa.

The ambient pressure R(T) data fits to an activated behaviour ( $\rho = \rho_0 \exp(\Delta / k_B T)$ in the 70 K to 300 K temperature range. The fitted curve is shown in Fig. 3 as solid red lines. Below 70 K the R(T) data fits to variable range hopping transport, probably arising from defect states in the semiconducting gap. Since we are interested in the electronic structure of the bulk we restrict to Arrhenius fits in the 70 K to 300 K temperature range for all pressures. The activation energies obtained from fits of the R(T) data as a function of external pressure are shown in the Fig.4b. It is evident from the figure that the activation energy shows a systematic decrease with increase in applied pressure. In particular, it decreases from from 38.4 $\pm$ 2meV at ambient pressure to to 3$\pm$ 0.2 meV at 4.3 GPa after which the crystal metallises. The magnitude of the ambient pressure, semi conducting gap obtained from our data is comparable to that obtained from an earlier report [7].

In addition to the observation of metallization by the application of pressure at 5.8 GPa, a notable fall in resistance was observed at a temperatures ~17 K, which is visible in all graphs in the right panel of Fig.3. The magnitude of the fall in resistance increases with increase of pressure and it is more clearly seen in the R(T) measured at the highest applied pressure of 8.2 GPa. The fall in resistivity is shown in expanded form in the inset of Fig3g. It is clear from the figure that there is a current-dependent broadening, in the onset of the fall in the resistivity. Such a current dependent shift was seen to occur in single crystals of $BaFe_2As_2$ at ~3.5 GPa, preceding the stabilisation of a more robust superconducting state at ~6.4 GPa, where no current dependence was seen [17].Our results shown in the inset of Fig.3g, seen in the light of that in ref [17] could imply that superconductivity occurs in a small fraction of the sample at ~8 GPa and that superconductivity could be more robust at higher pressures. Unfortunately, the experiments could not be done at higher pressure to provide an unambiguous evidence for a superconducting transition on account of the limitation of the pressure cell. High pressure low temperature magnetization experiments, to confirm diamagnetism associated with the superconducting like anomaly seen from 5.8 GPa will be useful.

In the light of the observation of metallization of $BaMn_2As_2$ ( cf. Fig.3), we have carried out high pressure powder XRD measurements at room temperature, to see if the metallization is associated with a structural transition. The measured variation of the ‘a’ and ‘c’ lattice parameters with external pressure are shown in Fig.5a. While both the lattice parameters, ‘a’ and ‘c’ decrease monotonically with increasing pressure, there is a distinct change in the pressure variation of lattice parameters, with a cross over region at ~ 5 GPa. The lattice parameter ‘a’ becomes less compressible at higher pressures, whereas, the compressibility of the lattice parameter ‘c’ shows a slight increase. The cube of the individual lattice parameters could be fitted to a third order Birch – Murnaghan (BM) Equation of State (EOS) and the zero pressure compressibilities ($\beta_{a0}$) of individual lattice parameters are indicated in the fig.5. The structural parameter of importance in the arsenide system is the c/a parameter [18], and Fig. 5b shows the variation of the c/a ratio as a function of pressure. The magnitude of c/a and its smooth variation with pressure indicates that the metallization observed in the present study is not associated with the collapsed tetragonal phase. Figure 5c shows the lattice volume versus pressure, and the fits to the third order Birch- Murnaghan equation of state. The difference in the compressibility behaviour for the low pressure insulating phase and high pressure metallic phase can be clearly seen, with a cross over at 5 GPa. The significance of the above pressure data can also be appreciated in the context of efforts [10] to metallise $BaMn_2As_2$ through substitutions.

The room temperature cell volume [ 1] in $BaT_2As_2$ plotted for T a transition 3d element from Cr to Cu reveals that the cell volume shows an anomalous increase for T=Mn, which is linked to the high spin state of Mn. It is noteworthy that the cell volumes observed upon metallization of $BaMn_2As_2$ is very close to the cell volumes of metallic $BaCr_2As_2$ and $BaFe_2As_2$, both of which are itinerant anti-ferromagnets. These observations suggest that cell volume plays a crucial role in the stabilization of metallic state in this system.

The insulator to metal transition( cf. Fig.3), that is associated with an iso-structural transition

(cf. Fig.5) points to an electronic topological transition induced by pressure. Iso-structural transitions have been observed in many insulating and semi-conducting systems where volume anomalies correspond to electronic topological transitions (ETT) involving large variation in the density of states at Fermi level [19]. The occurrence of ETT has also been invoked in the arsenide family [20]. Band structure calculations would be worthwhile to elucidate the nature of electronic transition, as also studies on the evolution of the magnetic state of $BaMn_2As_2$ as it metallizes under the application of pressure.

In conclusion, we have demonstrated metallization of the G-type antiferromagnetic $BaMn_2As_2$ under a application of pressure of ~5 GPa. High pressure XRD measurements at room temperature show that this insulator to metal transition is associated with an iso-structural electronic transition. The observed pressure induced metallization of $BaMn_2As_2$ , would help in fine tuning the on-going efforts on metallization with chemical substitutions. The metallic phase has a sharp drop in resistance at 17 K, suggestive of a superconducting transition.

The authors acknowledge Dr. Shemima Hussain and Dr. G. Amarendra of UGC DAE-CSR, Kalpakkam, node for the SEM characterization.

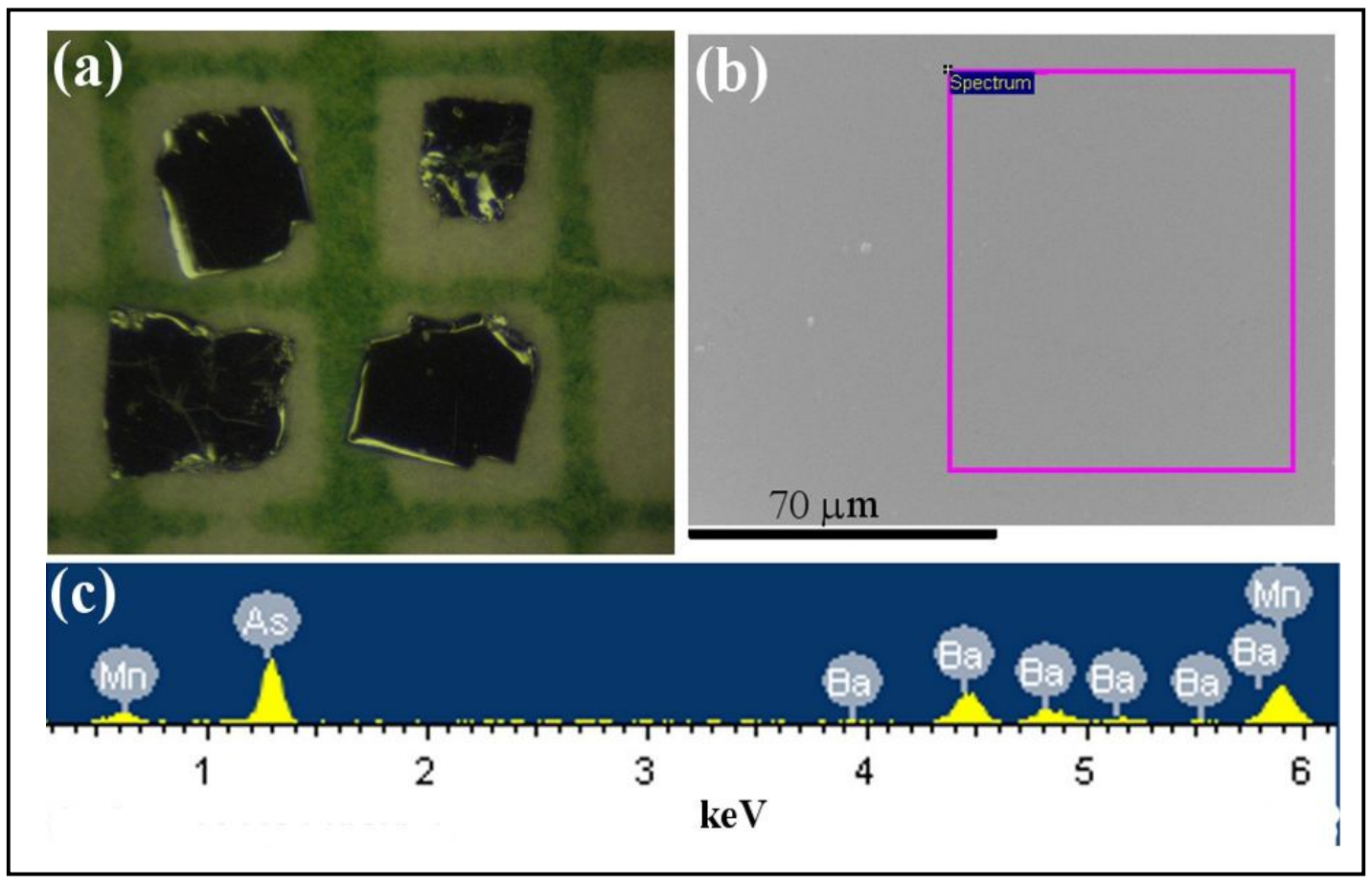

Fig.1 (a) photographs of the $BaMn_2As_2$ crystals, (b) shows a small region of one of the crystals, in which the energy dispersive x-ray pattern shown in (c) was obtained.

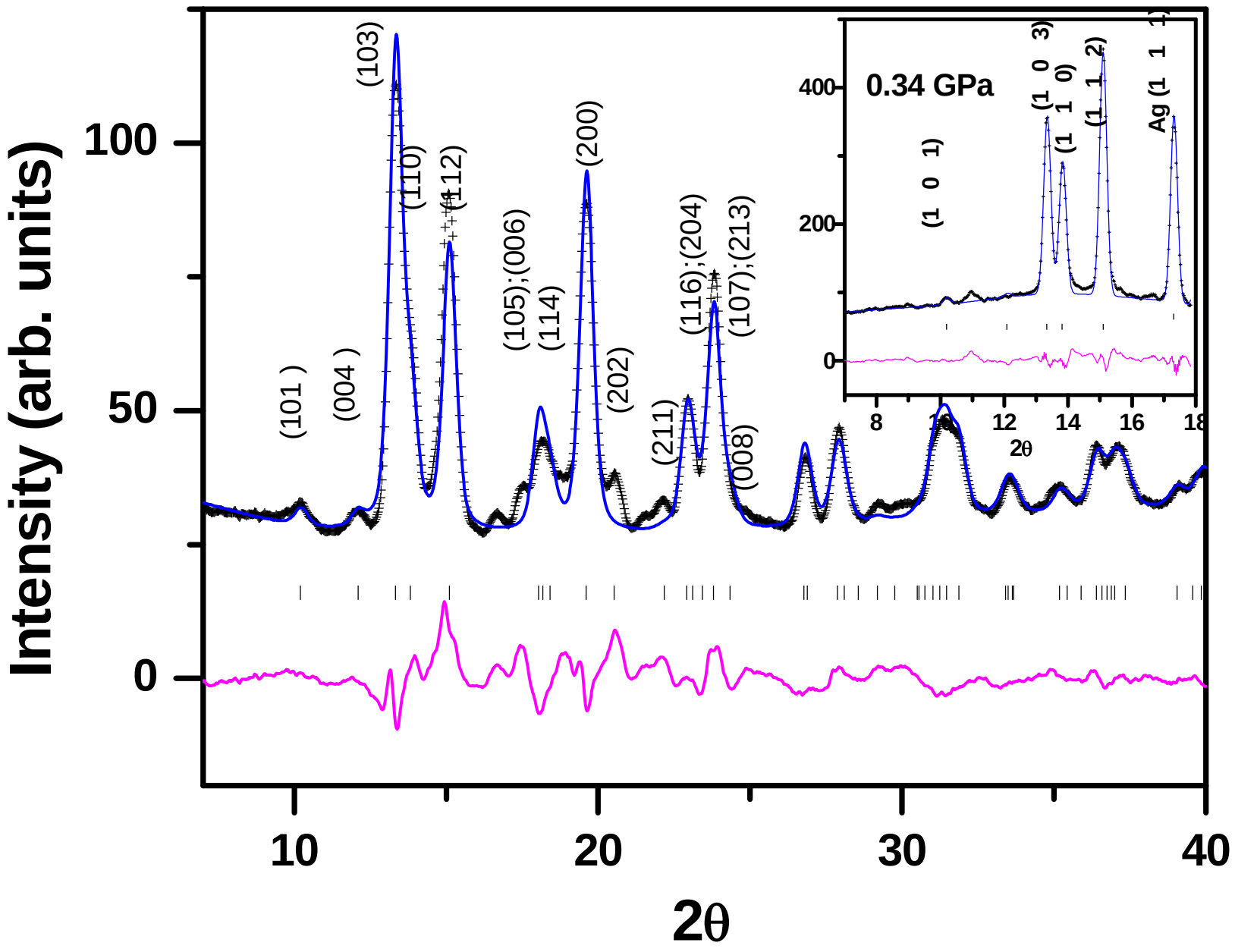

Figure.2 XRD powder diffraction pattern at ambient pressure obtained by powdering tiny $BaMn_2As_2$ single crystals embedded in the melt, using Mo X-ray beam and a Rigaku ULTRAX (18kW) rotating anode X-ray generator, monochromatised with graphite. The plot shows intensity versus 2θ values, of the observed (black +) and the calculated (blue solid line), obtained from Rietveld refinement. The difference (pink solid line) is also shown. In the inset the XRD data obtained in the DAC at 0.34 GPa is shown, along with the that of the pressure calibrant Ag (111) line. Data such as in the inset was used to obtain the lattice parameters shown in Fig.5.

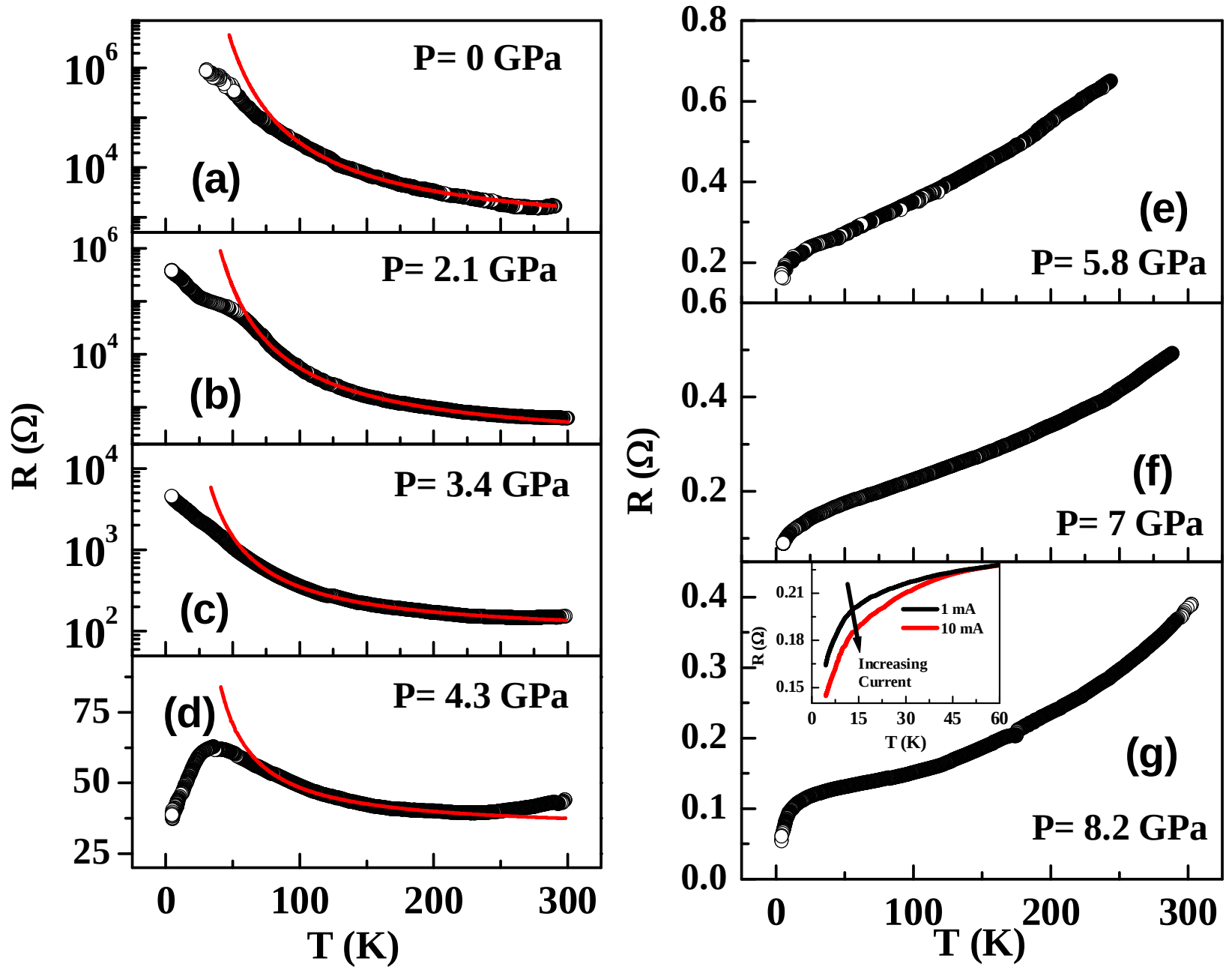

Fig. 3: Resistance versus temperature for $BaMn_2As_2$ at different pressures between 0 to 8.2GPa for representative pressures. Left panels indicate the pressure regime in which the sample shows an semiconducting behaviour. Red solid lines in Fig.3a to 3d are fits to $\rho = \rho_0 \exp(\Delta / k_B T)$ in the 70 K to 300 K temperature range. The right panel shows R(T) in the pressure regime in which the sample shows metallicity. Inset of Fig.3g shows the current dependence of the fall in resistivity at 17 K.

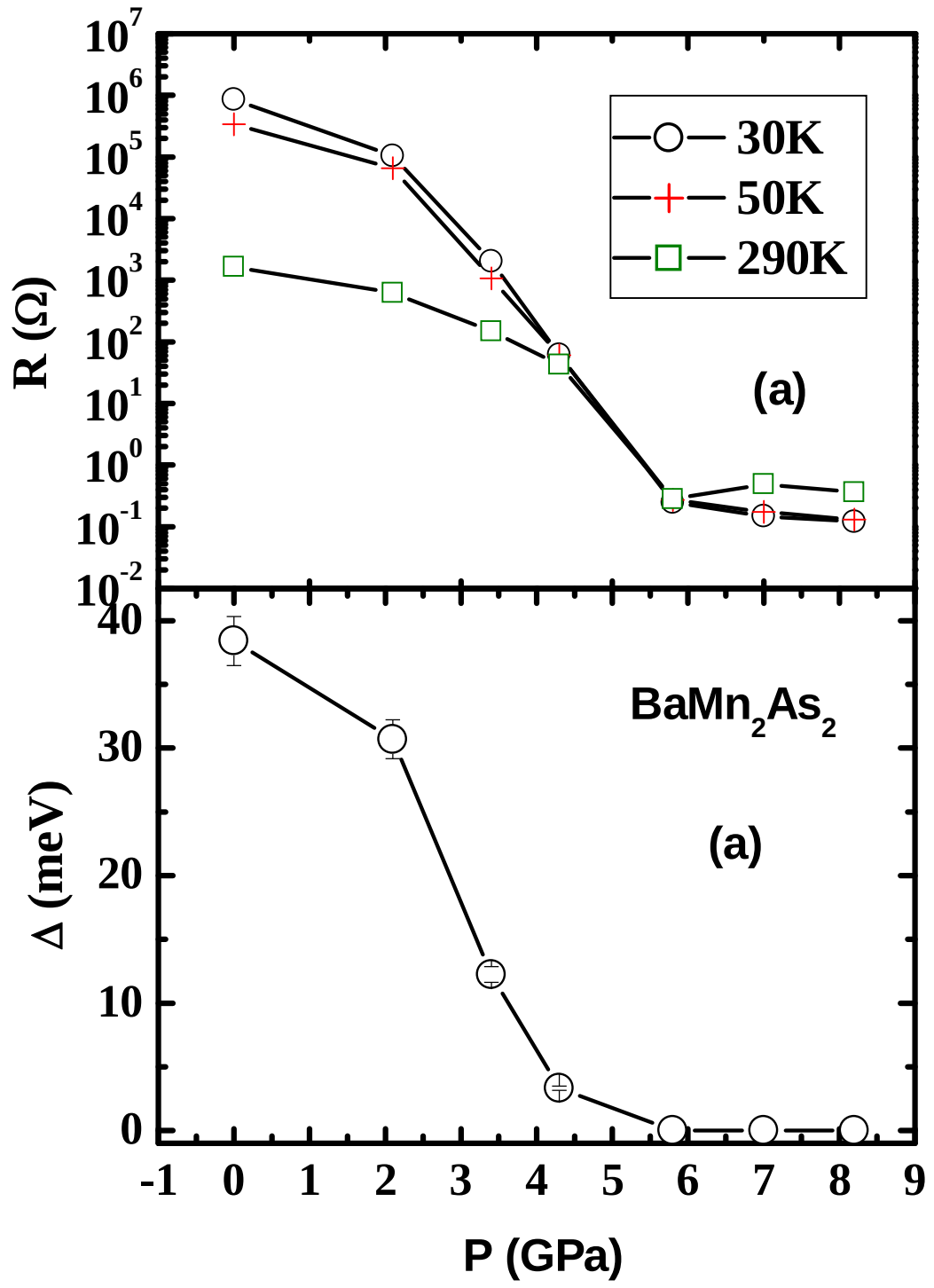

Fig.4 (a) Change in resistivity with temperature at various pressures to highlight the transition from insulating and metallic behavior at high pressures. (b) The band gap as a function of pressure determined from the Arrhenius fits of the R(T) data in the 70 K to 300 K temperature range.

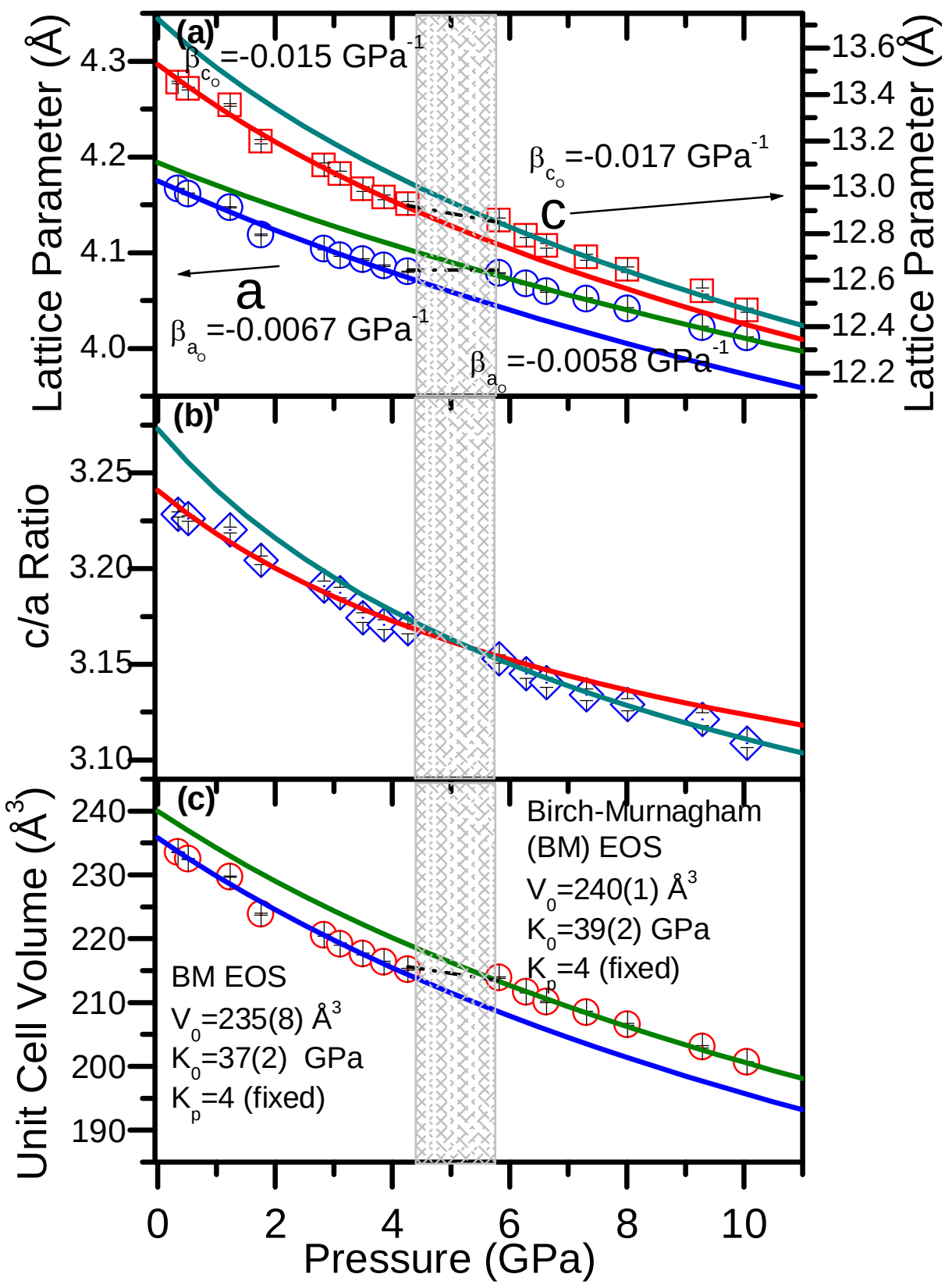

Fig.5 (a) The tetragonal lattice cell parameters of $BaMn_2As_2$, 'a' and 'c', versus applied pressure. (b) The c/a ratio versus pressure plot ( c) Unit cell volume versus pressure data. The solid lines are the third order fits to the Birch-Murnaghan equation of state (see text for details). The refined values of the zero pressure volumes as well as the bulk modulus are shown in the figure. The dotted line in the figures (a) and (c) are guides to the eye.